\documentstyle{aipproc2}

\input epsf

\begin{document}
\title{The 4 Year COBE DMR data is non-Gaussian}

\author{Pedro G. Ferreira$^{1,2}$, Krzysztof M. G\'orski$^{3,4}$, 
Jo\~ao Magueijo$^{5}$}
\address{
$^1$ Theory Group, CERN, CH-1211, Geneve 23 ,Switzerland\\
$^2$ CENTRA, Instituto Superior T\'ecnico, Lisboa 1096 Codex, Portugal\\
$^3$Theoretical 
Astrophysics Center, Juliane Maries Vej 30,
DK-2100, Copenhagen \O, Denmark\\
$^4$ Warsaw University Observatory, Warsaw, Poland\\
$^5$Theoretical Physics, 
Imperial College, Prince Consort Road, London SW7 2BZ, UK} 


\maketitle

\begin{abstract} 
I review our recent claim that there is evidence of non-Gaussianity
in the 4 Year COBE DMR data. I describe the statistic we apply,
the result we obtain and make a detailed list of the systematics
we have analysed. I finish with a qualitative understanding of what it might 
be and its implications.
\end{abstract}

\section{Introduction}

In a recent letter \cite{apjl1} we have claimed that the 4 Year COBE DMR data
exhibits evidence of non Gaussianity; our quantitative claim was that
the hypotheses that this data set is due to a Gaussian random process
can be ruled out at the $98 \%$ confidence level. This came somewhat as
a surprise given that this data set was lauded as strong evidence
for the inflationary paradigm: its statistics were thought
to be {\it consistent} with Gaussianity \cite{kog96a}.

We have performed an extensive analysis of the 4 year COBE DMR data set
and have been unable to find a non cosmological origin for the
non-Gaussian signal. The resuls of this analysis are presented in
\cite{bigpaper}. In this report I will only 
address a few often raised questions:
\begin{itemize}
\item Why does this statistic get a different result from all the
ones that were previously used?
\item Isn't the effect we are seeing just some systematic effect
of the analysis?
\item If it is signal what could it be?
\end{itemize}

\section{The Statistic and Result}
In our analysis we propose, and work with, an estimator for
 the {\it normalized bispectrum}.
To construct such an estimator we work
in the spherical harmonic representation $a_{\ell m}$ and
consider the tensor product of $3$ $\Delta T_\ell=\sum_{m}a_{\ell m}$s
One is interested in rotationally invariant quantities. These can be
trivially obtained if one rewrites the tensor product in terms of
the total angular momentum basis. The coefficient of the singlet will
be the higher order invariant we are looking for.  This procedure
leads to an estimator of the bispectrum
\begin{eqnarray}
{\hat B}_\ell&=&\alpha_\ell\sum_{m_1m_2m_3}{\cal W}^{\ell \ell
\ell}_{m_1m_2m_3} a_{\ell m_1}a_{\ell m_2} a_{\ell m_3}
\nonumber \\
\alpha_\ell&=&\frac{1}{(2\ell+1)^{\frac{3}{2}}}
\left ({\cal W}^{\ell \ell \ell}_{000}
\right )^{-1}
\label{bispec}
\end{eqnarray}
where ${\cal W}^{\ell_1\ell_2\ell_3}_{m_1m_2m_3}$ are the Wigner 3J
coefficients.  ${\hat B}_\ell$ may then 
be divided by the appropriate power of an estimator for $C_\ell$ in order to make 
it dimensionless, and suitably normalised.
\begin{eqnarray}
I^3_\ell &=&\left| { {\hat B}_\ell\over ({\hat C}_\ell)^{3/2}}
\right| \label{defI}
\end{eqnarray}
Let us comment on a number of features of the estimator. Firstly this 
statistic is global on pixel space, i.e. all the estimators are
a function of all the pixel values. This means that it will be very
good at identifying the scale dependence of non-Gaussianity (in the
same way as the $C_\ell$s are very discrimative of the scale dependence
of the variance of the fluctuations); however it performs very poorly
at identifying the location of any signal on the map. In this it
contrasts with statistics which have been applied to the COBE data
until now: the three point correlation function, the statistics of
peaks and topological measures are all defined on pixel space \cite{kog96a}.
Secondly we have assumed statistical isotropy in constructing this estimator.
We know, however, that the COBE data has a number of anisotropic
features which violate this assumption, in particular the presence
of the galaxy (which must be removed) and the anisotropic sky coverage
of the observation pattern. The only way we can truly assess the significance
of the $I^3_\ell$ is by comparing them to an ensemble of $I^3_\ell$
measured on Gaussian sky maps generated with exactly the same characteristics
as that of the 4 year COBE DMR data. The assumptions of the Monte Carlo
must be carefully checked. 

We have tested the  inverse noise variance weighted, average maps of 
the 53A, 53B, 90A and 90B {\it COBE}-DMR channels, with monopole
and dipole removed, at resolution 6, in ecliptic 
pixelization. We use the  
extended galactic cut of \cite{banday97}, and 
\cite{benn96} to remove most of the emission from the plane of the Galaxy.
We apply our statistics to the DMR maps before and after correction
for the plausible diffuse foreground emission outside the galactic plane
as described in
\cite{kog96b}, and \cite{COBE}. 
To  estimate the $I^3_\ell$s we set
the value of the pixels within the galactic cut to 0 and 
the average temperature {\it of the cut map} to zero. 
We then integrate the
map multiplied with spherical harmonics  to obtain the estimates of the
$a_{\ell m}$s and apply equations \ref{bispec} and \ref{defI}.
We then compare the estimates to the distributions generated from
Monte Carlos simulations of Gaussian maps.

\begin{figure}
\centering
\leavevmode\epsfysize=7cm \epsfbox{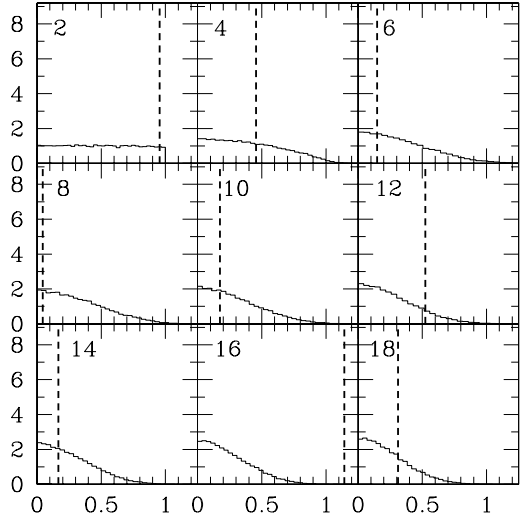}\\ 
\caption[flatness]{\label{fig1}The vertical thick dashed line represents the value 
of the observed
$I^3_\ell$.  The solid line is the probability distribution function
of $I^3_\ell$ for a Gaussian sky with extended galactic cut and
DMR noise.}
\end{figure}

\begin{figure}
\centering
\leavevmode\epsfysize=7cm \epsfbox{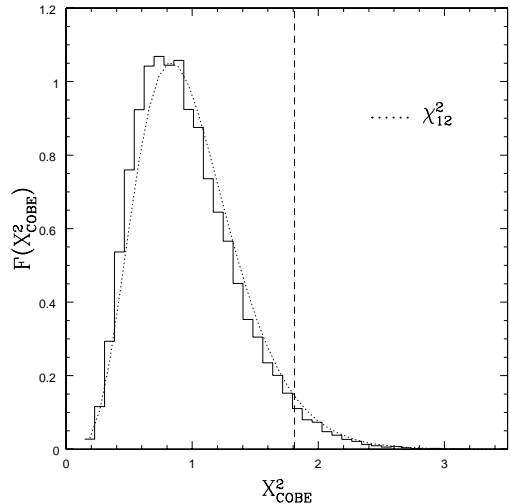}\\ 
\caption[chi2]{\label{chi2}The distribution $P(X^2_{COBE})$ and a fit to it, which 
is a chi squared with 12 degress of freedom. The observed chi
squared, with COBE, is represented as a vertical line: $X^2_{COBE}=1.81$. 
One can compute $P(X^2_{COBE}<1.81)= 0.98$.
Hence, it would appear that we can
reject Gaussianity at the $98\%$ confidence level.}
\end{figure}

Since the $I^3_\ell$ distributions are  non Gaussian  we 
generalize the $\chi^2$ for a set of probability functions
$P_\ell(I^3_\ell)$ associated with observations $\{I^3_\ell\}$ 
by defining the following functional
\begin{equation}\label{presc}
X^2={1\over N}{\sum_\ell X_\ell^2}=
{1\over N}{\sum_\ell (-2\log P_\ell(I^3_\ell) 
+ \beta_\ell),}
\end{equation}
where the constants $\beta_\ell$ are defined so that for each term
of the sum $\langle X_\ell^2\rangle=1$. The definition reduces
to the usual $X^2$ for Gaussian $P_\ell$. 
Again, we build a $X^2$ for the {\it COBE}-DMR data by means of Monte
Carlo simulations. We proceed as follows. First we compute the 
distributions $P(I^3_\ell)$, for $\ell=2,\dots,18$, 
for a Gaussian process as measured subject to our galactic
cut, and pixel noises. These $P(I^3_\ell)$ were inferred 
from 25000 realizations (see Fig.~\ref{fig1}). 
From these distributions we then build 
the $X^2$ defined in ($\ref{presc}$), 
taking special care with the numerical
evaluation of the constants $\beta_\ell$. We call 
this function $X^2_{COBE}$.
We then find its distribution $F(X^2_{COBE})$
from 10000 random realizations.  This is very well approximated by 
a $\chi^2$ distribution with 12 degrees of freedom
(Fig.~\ref{chi2}). 
We then compute $X^2_{COBE}$ with the actual observations and find
$X^2_{COBE}=1.81$. One can compute $P(X^2_{COBE}<1.81)= 0.98$.
Hence, it would appear that we can
reject Gaussianity at the $98\%$ confidence level.

\section{Is it a systematic effect?}

The non-Gaussianity we are finding in the 4 year COBE
DMR maps is surprising and fascinating enough that we have gone through
an exhaustive test of all possible systematic effects. We summarize them
in the following list:
\begin{quote}
\begin{enumerate}
\item Foregrounds contamination:
\begin{itemize}
\item Dust (using the DIRBE sky maps and also the Schlegel {\it et al}
dust model)
\item Synchrotron (with the Haslam template)
\item Foreground corrected maps
\end{itemize} 
\item Noise model:
\begin{itemize}
\item Anisotropic sky coverage
\item Noise correlations between different pixels 
\item Analysis of noise templates
\end{itemize}
\item Galactic cut:
\begin{itemize}
\item Dependence on shape (``custom'' versus constant elevation)
\item Dependence on elevation
\item Dependence on monopole and dipole subtraction, before or
after the cut, with or with out galaxy.
\end{itemize}
\item  Systematic templates 
\begin{itemize}
\item Spurious offsets induced by the cut.
\item Instrument susceptibility to the Earth 
magnetic field.
\item Callibration errors .
\item Errors due to incorrect removal of the COBE Doppler and  
Earth Doppler signals.
\item Errors in correcting for  emissions from the Earth, and
eclipse effects.
\item Artifacts due to uncertainty in the correction for
the correlation created by the low-pass filter on the
lock-in amplifiers (LIA) on each radiometer
\item Errors due to emissions from the moon, and the planets. 
\end{itemize}
\item Assumptions in Monte Carlos:
\begin{itemize}
\item Dependence on tilt
\item Dependence on smooth versus discontinuous power spectrum
\item Dependence on beam shape
\item Dependence on pixelization.
\end{itemize}
\end{enumerate}
\end{quote}

\begin{figure}
\centering
\leavevmode\epsfysize=7cm \epsfbox{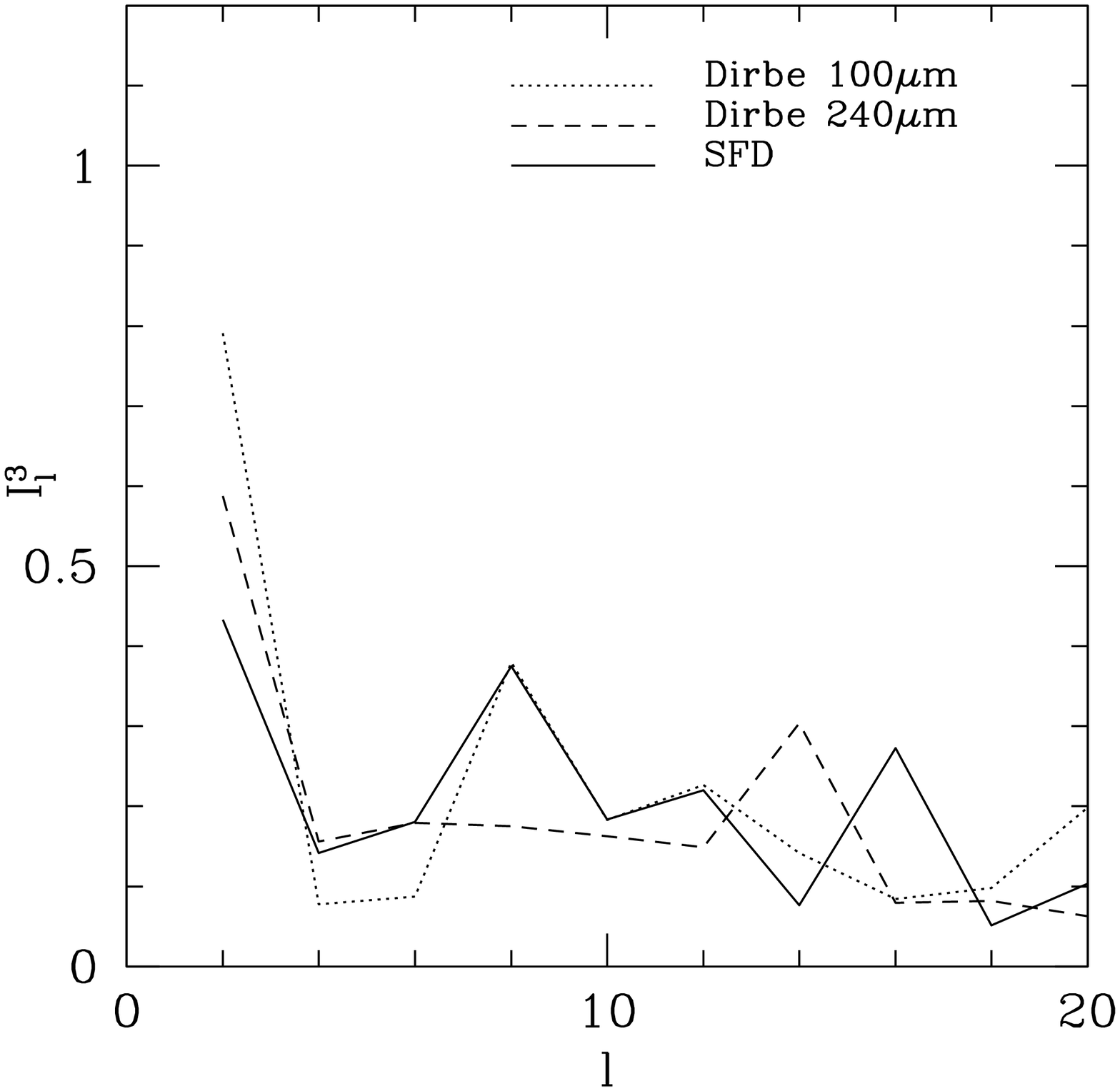}
\leavevmode\epsfysize=7cm \epsfbox{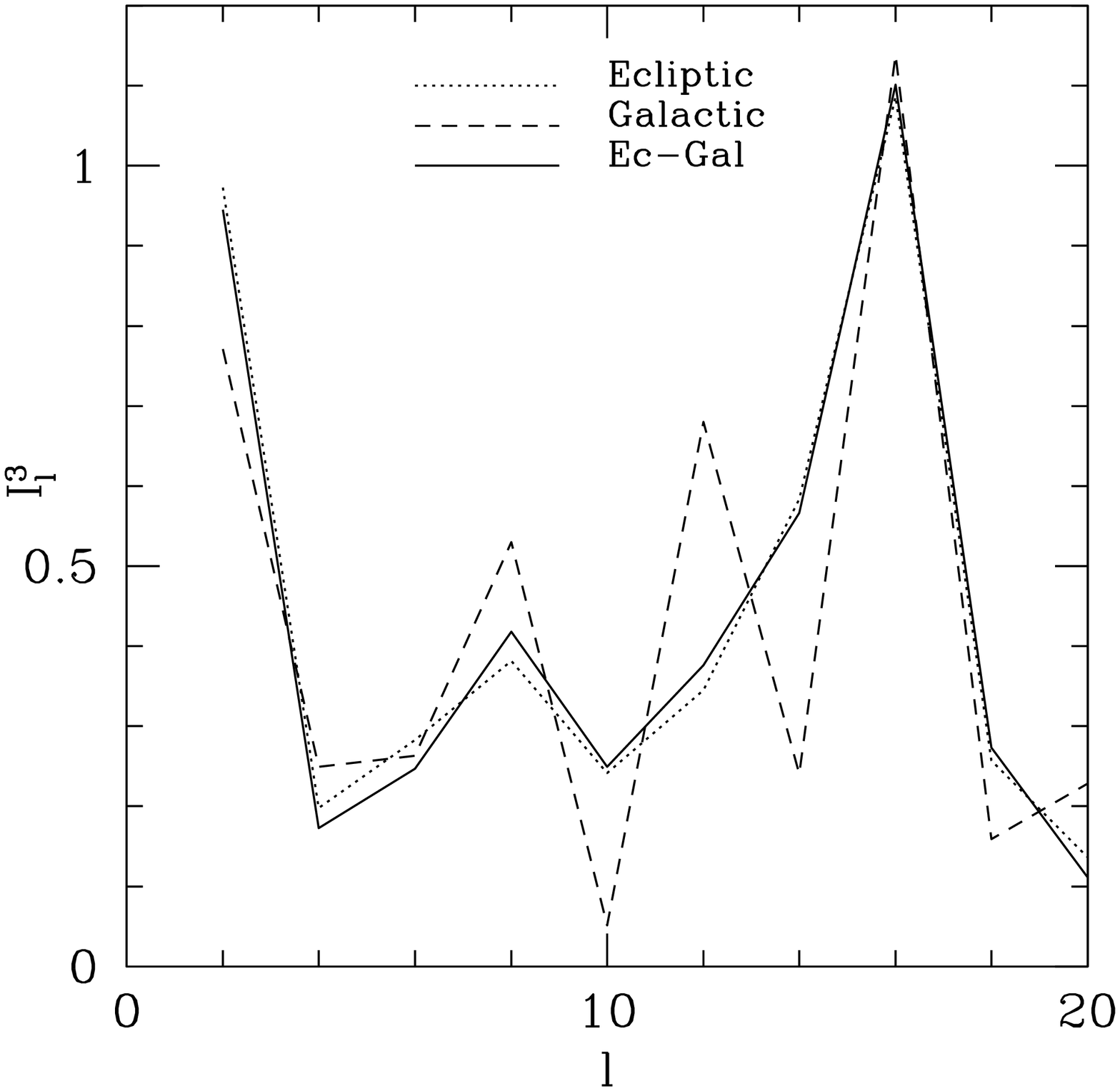}
\caption[foreg]{\label{foreg}Left panel: The measured $I_\ell$ for dust galactic foregrounds.
We have plotted values concerning the DIRBE 100$\mu$m and DIRBE 240$\mu$m 
as well as the Schlegel, Finkbeiner, and  Davis maps. Right panel: The measured $I_\ell$ for DMR
foreground corrected data in ecliptic coordinates, 
galactic coordinates, and ecliptic coordinates with
galactic frame coupling.}
\end{figure}

Let us just highlight the foreground tests. In figure \ref{foreg} (left panel)
 we plot
the $I^3_\ell$s estimated directly from the dust maps produced by the
DIRBE $100\mu m$ and $240\mu m$ channels and from the dust template
constructed in \cite{sfd98} using the DIRBE and IRAS maps. The feature
we identify in the DMR data is not there. Furthemore, if we use the
dust templates to subtract any foreground contribution (see
figure \ref{foreg} right panel) the value of $I^3_{16}$ actually increases!

We have found that none of the effects listed above change our result by much, in fact
for most corrections the confidence level rises to over $99\%$. We therefore
claim that the non-Gaussianity we find in the COBE DMR data is not due to
any of the {\it known} systematics.
\section{The nature of the signal in $I^3_{16}$}
The structure of the non-Gaussian signal is truly intriguing. It manifests
itself as a spectral ``spike'' at $\ell=16$; it is difficult to associate
such a pattern to some known or speculated source of fluctuations. 
If the signal is cosmological, the minimal inflationary models
cannot be right. On the other hand it is not obvious that 
the main competitor to inflation, topological defects, could 
explain this type of non-Gaussianity. Topological defects 
are non-Gaussian, but in ways which are often more
subtle than commonly thought. 
An interesting possibility was recently proposed by Peebles
\cite{peeb}. This is an isocurvature model in which the underlying
fluctuations are not a Gaussian random field, but the square 
of a Gaussian random field. The model is based on 
non minimal inflation, but produces fluctuations radically different
from minimal inflationary fluctuations. 

A more useful exercise is to try and understand, given the characteristics
of the experiment, what the power spectrum of such a non-Gaussian signal
might be. The simplest thing to consider is combination of three
signals. On very large scales (upto $\ell=12-14$) we have a Gaussian
sky signal. It may be fundamentally Gaussian or merely a manifestation
of the law of large numbers. For  $\ell\ge14$ the sky signal is non-Gaussian,
and this should manifest itself all the way upto much higher $\ell$s. However
the instrument noise of the 4 Year COBE DMR data sets starts to dominate the
$I^3_{\ell}$s at $\ell=18-20$, where the signal to noise drops below unity.
Given that the noise has been shown to be extremely well characterized by
a Gaussian, the sky map will manifest itself as Gaussian for these higher $\ell$s. 

Following they publications of our result, a number of groups
have reported similar results: Pando {\it et al} \cite{pando} have applied a
wavelet based technique and find evidence that the non-Gaussian
signal is localized in the northern hemisphere (a result we tentatively
confirm \cite{bigpaper}) while Novikov {\it et al} \cite{nov} have applied topological tests
to detect non-Gaussianity in the COBE DMR data. According to skeptics, 
this may merely reflect a change in the 
psychological prior, triggered by our work. More seriously
one should remember that the work performed by us and by these
groups makes use of the same data set. Therefore this work 
provides an independent confirmation of our analysis of the DMR
maps. In the very least this may mean a revision of the data
analysis techniques which are currently in vogue for power spectrum
estimation from CMB data sets: the underlying assumption in
these Bayesian techniques is that the data set is Gaussian
(greatly simplifying the estimation algorithms) \cite{gorski97}. However the
fact that COBE data set seems to be non-Gaussian  raises the
question weather the current estimates of the $C_\ell$s of
the COBE data set are as accurate as they are claimed to be.

If the sky is truly non-Gaussian in the way we describe above, then
we are at the threshold of uncovering its statistical nature with
the higher resolution experiments that are coming online in particular with 
the BOOMERANG LDB experiment (described elsewhere in these proceedings)
and ultimately with the MAP and Planck Surveyor satellite experiments.
\section*{Acknowledgments}

We thank the organizers for an excellent meeting. We thank
JNICT, NASA-ADP, NASA-COMBAT, NSF,  RS, Starlink, TAC for support.

\end{document}